\documentclass[review,12pt,times,preprint]{elsarticle}

\usepackage{graphicx}
\usepackage{bm}
\usepackage{amsmath}
\usepackage{lineno}
\usepackage{color}
\usepackage[left=2cm,right=2cm,top=2cm,bottom=2cm]{geometry}

\def\nuc#1#2{\relax\ifmmode{}^{#1}{\protect\text{#2}}\else${}^{#1}$#2\fi}

\journal{Physics Letters B}
\biboptions{comma,sort&compress}
\begin{document}
    
    \begin{frontmatter}
        
\title{Binding-energy independence of reduced single particle strengths derived from $(p,d)$ reactions}

        \author[label1]{Y.P.~Xu}
        \author[label1,label2]{D.Y.~Pang\corref{cor1}}
        \ead{dypang@buaa.edu.cn}
        \author[label1]{X.Y.~Yun}
        \author[label3]{C.~Wen}
        \author[label4]{C.X.~Yuan}
        \author[label3]{J.L.~Lou}
        \cortext[cor1]{Corresponding author}
        \address[label1]{School of Physics and Nuclear Energy Engineering, Beihang University,
            Beijing 100191, China}
        \address[label2]{Beijing Key Laboratory of Advanced Nuclear Materials and Physics, Beihang University, Beijing 100191, China}
        \address[label3]{School of Physics and State Key Laboratory of Nuclear Physics and Technology, Peking University, Beijing 100871, China}
        \address[label4]{Sino-French Institute of Nuclear Engineering and Technology, Sun Yat-Sen University, Zhuhai 519082, China}

\date{\today}

\begin{abstract}
An overall reduction factor (ORF) is introduced for studying the quenching of single particle strengths through nucleon transfer reactions. The ORF includes contributions of all the probed bound states of the residual nucleus in a transfer reaction and permits a proper comparison with results of inclusive knockout reactions. A systematic analysis is made with 103 sets of angular distribution data of $(p,d)$ reactions on 21 even-even targets with atomic mass numbers from 8 to 56 using the consistent three-body model reaction methodology proposed in [J. Lee, J.A. Tostevin, B.A. Brown, et al., Phys. Rev. C 73, 044608 (2006)]. The extracted ORFs are found to be nearly independent on the nuclear isospin asymmetry, which is different from the systematics of inclusive knockout reactions but is consistent with the recent measurement of $(d,t)$, $(d,\nuc{3}{He})$, $(p,2p)$, and $(p,pn)$ reactions on nitrogen and oxygen isotopes and \textit{ab initio} calculations.
\end{abstract}

\begin{keyword}
    (p,d) reactions \sep 
    spectroscopic factors \sep
    quenching factors \sep
    reduced single particle strength
\end{keyword}
\end{frontmatter}

\section{Introduction}

Quenching of single particle strength (SPS) is an important subject in nuclear physics studies \cite{Rohe-PRL-2004, Gade-PRL-2004, Gade-PRC-2008, Natasha-PRL-2009, Barbieri-PRL-2009, Jenny-PRL-2010, Kay-PRL-2013, Flavigny-PRL-2013, Tostevin-PRC-2014, Atar-PRL-2018, Moro-PLB-2018, Kawase-PTEP-2018,Flavigny-PRC-2018}. It was firstly observed in $(e,e'p)$ reactions on some stable nuclei \cite{Herder-NPA-1988, Lapikas-NPA-1993, Kramer-NPA-2001}. Spectroscopic factors (SFs) deduced from those experimental data are found to be 40-60\% lower than the sum-rule limit given by independent-particle shell model. Such quenching of SPS has been attributed to some profound questions in nuclear physics, such as short- and medium-range nucleon-nucleon correlations and long-range correlations from coupling of the single-particle motions of the nucleons near the Fermi surface and the collective excitations \cite{Herder-NPA-1988, Dickhoff-PPNP-2004, Pandharipande-RMP-1997}.

Systematic studies of nucleon (proton or neutron) knockout reactions performed with radioactive nuclei on light targets (Be and C) suggest that the quenching factors, or the reduction factors (RFs), $R_s$, of the SPS carry a strong dependence on the isospin asymmetry, $\Delta S$, of the projectile nuclei \cite{Gade-PRC-2008, Tostevin-PRC-2014}. $\Delta S$ is defined to be the difference between the neutron and proton separation energies ($S_n$ and $S_p$, respectively) of the particles concerned, i.e., $\Delta S=S_n-S_p$ for neutron removal and $\Delta S=S_p-S_n$ for proton removal. (In practice, effective $\Delta S$ values are defined, which take into account the excitation energies of the reaction residues \cite{Simpson-PRC-2009, Tostevin-PRC-2014}). The $R_s$ values deduced from knockout reactions are found to be close to unity when the removed nucleons are weakly-bound ($\Delta S \lesssim -20$ MeV) and are very small when the removed nucleons are strongly-bound ($\Delta S \gtrsim 20$ MeV). However, there is no strong evidence for such strong dependence in the $R_s$ values obtained from systematic studies of transfer reactions, such as $(p,d)$ and $(d,p)$ reactions \cite{Jenny-PRL-2010, Kay-PRL-2013} and $(d,t)$ and $(d,\nuc{3}{He})$ reactions \cite{Flavigny-PRL-2013, Flavigny-PRC-2018}. Such discrepancy also exists in structure theory. By solving the Pinkston-Satchler inhomogeneous equation with correlation-dependent effective nucleon-nucleon interactions, N.K. Timofeyuk found that strong isospin asymmetry dependence of the quenching factors exist with light exotic nuclei \cite{Timofeyuk-PRL-2009}. But such dependence was not found to exist in some ab-initio calculations \cite{Barbieri-IJMPA-2009, Cipollone-PRC-2015, Atar-PRL-2018}. Recently, the RFs are also found to be independent on the isospin asymmetry in $(d,t)$, $(d,\nuc{3}{He})$, $(p,2p)$, and $(p,pn)$ reactions on nitrogen and oxygen isotope \cite{Flavigny-PRL-2013, Atar-PRL-2018, Kawase-PTEP-2018, Moro-PLB-2018,Flavigny-PRC-2018}.

It is still an open question about why the dependence on the isospin asymmetry differ systematically between the RFs obtained from knockout and transfer reactions. One important thing to notice is that the $R_s$ values are defined differently in these two types of reactions. In the knockout reactions compiled in Ref. \cite{Tostevin-PRC-2014}, the experimental one-neutron removal cross sections are mostly inclusive, that is, contributions from all the excited states of the knockout residues below their particle emission thresholds were included in the measured data. Therefore, the total one-neutron removal cross sections should be calculated as sums of the single-particle cross sections \cite{Simpson-PRC-2009}:
\begin{align}
\label{eq-sigma-ko}
\sigma_{-1n}=\sum_{nlj}\left[\frac{A}{A-1}\right]^{N_{nl}} C^2S(J^\pi, nlj)\sigma_\textrm{sp}(nlj, S_n)
\end{align}
where $C^2S(J^\pi, nlj)$ are the shell model SFs which depend on the spin-parties of the core states, $J^{\pi}$, and the quantum numbers of the single particle 
states of the removed nucleon, $nlj$. The factors $[A/(A-1)]^{N_{nl}}$ are for the centre-of-mass corrections to the shell model SFs, where $N_{nl}$ is the number of the oscillator quanta associated with the major shell of the removed particle, which depends on the node number $n$ and orbital angular momentum $l$, and $A$ is the mass number of the composite nucleus \cite{Dierperink-PRC-1974, Tostevin-PRC-2014}. For sake of clarity, we designate $\left[\frac{A}{A-1}\right]^{N_{nl}} C^2S(J^\pi, nlj)$ by SF$^\textrm{th}$, which stands for a theoretical spectroscopic factor. The single particle cross sections, $\sigma_\textrm{sp}$, which depend on the quantum numbers ($nlj$) and the binding energies of the removed nucleons, include contributions from both the stripping and diffraction dissociation mechanisms \cite{Tostevin-NPA-2001} and are calculated using the eikonal model assuming unit SFs. The RFs in knockout reactions, $R_s^\textrm{ko}$, are defined as the ratio between the experimental ($\sigma_{-1n}^\textrm{exp}$) and theoretical one-neutron removal cross sections ($\sigma_{-1n}^\textrm{th}$):
\begin{equation}
\label{eq-rsko}
R_s^\textrm{ko}=\frac{\sigma_{-1n}^\textrm{exp}}{\sigma_{-1n}^\textrm{th}}
=\frac{\sigma_{-1n}^\textrm{exp}}{\sum_{nlj}\textrm{SF}^\textrm{th}(J^\pi, nlj)\sigma_\textrm{sp}(nlj, S_N)}
\end{equation}

In a transfer reaction, the reduction factor $R_s^\textrm{tr}$ is used to be defined as the ratio between the experimental and theoretical SFs \cite{Betty-PRL-2005, Jenny-PRC-2006, Betty-PRL-2009, Jenny-PRL-2010}:
\begin{equation}
\label{eq-rstr-1}
R_s^\textrm{tr}=\frac{\textrm{SF}^\textrm{exp}(J^\pi, nlj)}{\textrm{SF}^\textrm{th}(J^\pi, nlj)}.
\end{equation}
The experimental spectroscopic factor, $\textrm{SF}^\textrm{exp}$, is obtained by matching the experimental differential cross sections by the theoretical ones, usually at the angles where the experimental cross sections are on the maximum:
\begin{equation}\label{eq-sftr-1}
\textrm{SF}^\textrm{exp}=\left(\frac{d\sigma}{d\Omega}\right)^\textrm{exp}/\left(\frac{d\sigma}{d\Omega}\right)^\textrm{th}.
\end{equation}
The theoretical transfer cross sections are also calculated assuming the SFs being unity. Inserting Eq. (\ref{eq-sftr-1}) in Eq. (\ref{eq-rstr-1}), one gets the RF associated with a specific channel $(J^\pi,nlj)$ of transfer reaction:
\begin{equation}\label{eq-rstr-2}
R_s^\textrm{tr}(J^\pi,nlj)=\frac{\left[\frac{d\sigma}{d\Omega}(J^\pi,nlj)\right]^\textrm{exp}}{\textrm{SF}^\textrm{th}\left[\frac{d\sigma}{d\Omega}(J^\pi,nlj)\right]^\textrm{th}}.
\end{equation}

Comparisons between the so-defined RFs in knockout and transfer reactions, in Eqs. (\ref{eq-rsko}) and (\ref{eq-rstr-2}), respectively, have been made in, \textit{e.g.}, Refs. \cite{Jenny-PRC-2006, Jenny-PRL-2010, Flavigny-PRL-2013, Flavigny-PRC-2018}. However, the difference in Eqs. (\ref{eq-rsko}) and (\ref{eq-rstr-2}) is obvious. The reduction factor defined in Eq. (\ref{eq-rsko}) for an inclusive knockout reaction corresponds to, in principle, all the bound states of the knockout residue. Such RFs represent the \textit{averaged} deviation of the actual SPSs from the theoretical ones. On the other hand, the reduction factor defined in Eq. (\ref{eq-rstr-2}) corresponds to only one state (usually being the ground state) of the residual nucleus. One may argue that the two definitions of RFs correspond to different quantities and should not be compared directly.

A proper comparison between the RFs from transfer and knockout reactions may be made by assuming the transfer cross sections to be measured inclusively as well. In such a case, similar to Eq. (\ref{eq-rsko}), one can define an \textit{overall} reduction factor (ORF) for a transfer reaction:
\begin{equation}\label{eq-rstr}
R_s^\textrm{tr}=\frac{\sum_i\left(\frac{d\sigma}{d\Omega}\right)^\textrm{exp}_i}{\sum_i \textrm{SF}_i^\textrm{th}\left(\frac{d\sigma}{d\Omega}\right)^\textrm{th}_i}
=\frac{\sum_i \textrm{SF}_i^\textrm{exp}\left(\frac{d\sigma}{d\Omega}\right)^\textrm{th}_i}{\sum_i \textrm{SF}_i^\textrm{th}\left(\frac{d\sigma}{d\Omega}\right)^\textrm{th}_i}.
\end{equation}
The sums run over all the measured states of the residual nucleus. If we define a coefficient $A_i$ for each state by:
\begin{equation}\label{eq-Ai}
A_i= \left(\frac{d\sigma}{d\Omega}\right)^\textrm{th}_i/\left[\sum_i \textrm{SF}_i^\textrm{th}\left(\frac{d\sigma}{d\Omega}\right)^\textrm{th}_i\right],
\end{equation}
and assign an uncertainty for each $\textrm{SF}_i^\textrm{exp}$ by $\Delta_{ \textrm{SF}_i^\textrm{exp}}$, the uncertainty in $R_s^\textrm{tr}$, which is now $R_s^\textrm{tr}=\sum_i \textrm{SF}_i^\textrm{exp} A_i$, can be expressed as:
\begin{equation}\label{eq-rs-err}
\Delta R_s^\textrm{tr} = \sqrt{\sum_i A_i^2\Delta^2_{\textrm{SF}_i^\textrm{exp}}}.
\end{equation}
As usual, $R_s$ in Eq. (\ref{eq-rstr}) and $A_i$ in Eq. (\ref{eq-Ai}) are evaluated at the peaks of the angular distributions.

In order to see how the ORFs defined in Eq. (\ref{eq-rstr}) depend on the isospin asymmetry, we analyze 103 sets of angular distributions of $(p,d)$ reactions on 21 even-even nuclei, namely, \nuc{8}{He}, \nuc{12,14}{C}, \nuc{14,16,18}{O}, \nuc{22}{Ne}, \nuc{26}{Mg}, \nuc{28,30}{Si}, \nuc{34}{S}, \nuc{34,36,38}{Ar}, \nuc{40,42,44,48}{Ca}, \nuc{46}{Ar}, \nuc{46}{Ti}, and \nuc{56}{Ni}. The choice of target nuclei are mainly limited by the availability of experimental data. Even-even targets are chosen by practical reasons. In $(p,d)$ reactions with an even-even target, which has nil spin, there is only one single particle state corresponds to a state of the residue, which makes the theoretical analysis much easier than for $(p,d)$ reactions on a target with nonzero spin. Our analysis take into account several ($2\sim4$) bound states of the residual nuclei except for the \nuc{8}{He}$(p,d)$\nuc{7}{He} and \nuc{14}{O}$(p,d)$\nuc{13}{O} reactions, which have only ground state data available. The $\Delta S$ values range from $-22.3$ MeV to $18.6$ MeV for these reactions.

\section{Data Analysis}

It is well-known that reaction theories have been very successful in describing the angular distributions of transfer cross sections. But the amplitudes of these cross sections suffer from considerable uncertainties and the resulting SFs can be uncertain by around 30\% even if the statistical errors of the experimental data are reported small \cite{Betty-PRL-2005, Jenny-PRC-2006, Betty-arxiv, Betty-PRL-2009}. The reasons of this inaccuracy are typically attributed to the ambiguities in the entrance- and exit-channel optical model potentials (OMPs) and the single-particle potential (SPP) parameters. In view of such problems, the authors in Ref. \cite{Jenny-PRC-2006} proposed a consistent three-body model reaction methodology (TBMRM) for the analysis of $(p,d)$ and $(d,p)$ reactions. Such a methodology consists of adopting the Johnson-Soper adiabatic approximation for $(p,d)$/$(d,p)$ reactions \cite{Johnson-Soper}, of constraining the SPP parameters using modern Hartree-Fock calculations, and of calculating the nuclon-target OMPs by folding the effective JLM nucleon-nucleon interaction \cite{JLM-PRC-1977} with the nucleon density distributions from the same Hartree-Fock calculations. The deduced neutron SFs from $(p,d)$ and $(d,p)$ reactions on nuclei ranging from B to Ti with TBMRM are found to be suppressed by about 30\% with respect to large-basis shell model expectations and are consistent with the results of intermediate-energy nucleon knockout reactions within a limited range of $\Delta S$ values \cite{Jenny-PRC-2006}. 

We adopt the consistent TBMRM in this work. The details of this methodology can be found in Ref. \cite{Jenny-PRC-2006}. We hereby only briefly describe how it is applied in this work. With the Johnson-Soper adiabatic approximation, we only need OMPs for the $p$-$A$, $p$-$B$, and $n$-$B$ systems in a $A(p,d)B$ reaction. These potentials are obtained by folding the effective JLM nucleon-nucleon interaction with nucleon density distributions given by Hartree-Fock calculations. The real and imaginary parts of these nucleon OMPs are scaled with the conventional factors $\lambda_V=1.0$ and $\lambda_W=0.8$ \cite{Petler-PRC-1985, Jenny-PRC-2006}. The $p$-$B$ and $n$-$B$ potentials are evaluated at half energy of the deuteron in the exit channel. The consistent TBMRM adopts the same procedure as that used in the systematic analysis of knockout reactions \cite{Gade-PRC-2008, Tostevin-PRC-2014} for determining the geometry parameters, $r_0$ and $a_0$, of conventional Woods-Saxon potentials that generate the neutron single-particle wave functions, or overlap functions. With such a procedure, the diffuseness $a_0$ is fixed to be 0.7 fm. The radius parameter $r_0$ is adjusted so that the mean square radius of the transferred neutron orbital is $\langle r^2 \rangle = [A/(A-1)]\langle r^2\rangle_\textrm{HF}$, where $\langle r^2\rangle_\textrm{HF}$ is the value given by HF calculations and $A$ is the mass number of the composite nucleus. This adjustment is carried out with the separation energy given by HF calculation. The factor $[A/(A-1)]$ corrects the fixed potential center assumption used in the HF calculations. For all the cases studied in this work, the HF calculations are made with the Skyrme SkX interaction \cite{Brown-PRC-1998}, which is the same as those adopted in analysis of transfer and  knockout reactions, e.g., Refs. \cite{Gade-PRL-2004, Jenny-PRC-2006, Gade-PRC-2008, Jenny-PRL-2010, Tostevin-PRC-2014}. Once $r_0$ and $a_0$ are determined, the depths of the single-particle potentials are determined using the separation energy prescription with \textit{experimental} separation energies.

With the nucleon-nucleus potentials and neutron SPP parameters determined, we apply the Johnson-Soper adiabatic approximation in $(p,d)$ reaction calculations. All calculations are made with the computer code TWOFNR \cite{Tostevin-twofnr}. Information of these reactions are listed in the Supplemented Material, which include the target nuclei, the incident energies, the excited states of the residue, the $nlj$ values of the transferred neutrons, the $r_0$ values confined by HF calculations, the experimental and shell model SFs, and the resulting ORFs.

\begin{figure}[htbp]
    \centering
    \includegraphics[width=0.48\textwidth]{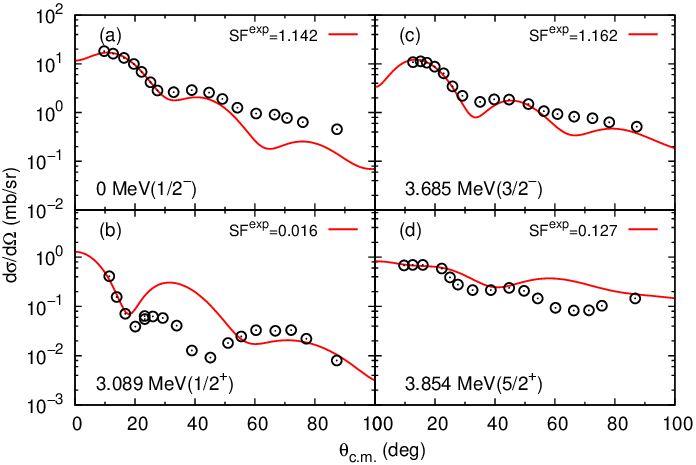}
    \caption{Comparisons between theoretical (curves) and experimental (circles) angular distributions for the \nuc{14}{C}$(p,d)$\nuc{13}{C} reaction at an incident energy of 35 MeV. Theoretical results are normalized to the experimental data with the spectroscopic factors indicated in the figures for the 0.0 MeV (a), 3.089 MeV (b), 3.685 MeV (c), and 3.854 MeV (d) states of \nuc{13}{C}.}
    \label{fig-fig1}
\end{figure}

As an example, we show the analysis of the \nuc{14}{C}$(p,d)$\nuc{13}{C} reaction at an incident energy of 35 MeV \cite{Yasue-NPA-1990}. The reaction residue \nuc{13}{C} has only four bound states below its particle emission threshold, which is 4.946 MeV for neutron decay. The angular distributions corresponding to these states are depicted in Fig. \ref{fig-fig1} together with the theoretical ones, which are normalized to the former at the peaks of these angular distributions, from which we obtained the experimental SFs. One sees that the calculations reasonably reproduced the two negative-parity state data, namely, the ground- and the 3.685 MeV states with $J^\pi={\frac{1}{2}}^-$ and ${\frac{3}{2}}^-$, respectively. Data of the other two states are not reproduced as satisfactorily but they are close to those reported in the original paper \cite{Yasue-NPA-1990}. By adopting the three-body model reaction methodology \cite{Jenny-PRC-2006}, which defined all quantities used in $(d,p)$/$(p,d)$ reaction calculations without free parameters, we do not attempt to improve the reproduction to these data by adjusting any reaction calculation parameters. On the other hand, as we will see below, the contributions to the ORF from these two positive-parity states are negligible. The details of this reaction are listed in Table. \ref{tab-14C-sfs} together with the SFs from shell model calculations with the YSOX interaction \cite{Yuan-PRC-2012}. The uncertainties of the extracted SFs, which contains both experimental and theoretical uncertainties, are difficult to evaluate. We adopt the global uncertainty of 20\% deduced in the systematic analysis of $(d,p)$/$(p,d)$ reactions by Tsang et al. \cite{Betty-arxiv} in this work.

\begin{table}[htbp]
    \centering
    \caption{Spectroscopic factors ($\textrm{SF}^\textrm{exp}$) extracted from the \nuc{14}{C}$(p,d)$\nuc{13}{C} reaction at an incident energy of 35 MeV. Listed are the excitations energies of \nuc{13}{C} ($E_\textrm{ex}$), their corresponding single-particle states ($nlj$), the HF confined $r_0$ values and shell model predicted SFs ($\textrm{SF}^\textrm{th}$).}
    \begin{tabular}{ccccc}
        \hline
        $E_\textrm{ex}$ (MeV)   & $nlj$   & $r_0$ (fm)  & $\textrm{SF}^\textrm{exp}$ & $\textrm{SF}^\textrm{th}$ \\\hline
        0.0    & $0p_{3/2}$ & 1.344  & 1.142  & 1.607  \\
        3.089 & $1s_{1/2}$ & 1.250  & 0.016  & 0.024  \\
        3.685 & $0p_{3/2}$ & 1.299  & 1.162  & 2.207  \\
        3.854 & $1d_{5/2}$ & 1.159  & 0.127  & 0.114  \\\hline
    \end{tabular}%
    \label{tab-14C-sfs}%
\end{table}%

\begin{figure}[htbp]
    \centering
    \includegraphics[width=0.48\textwidth]{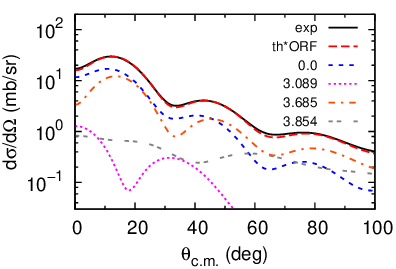}
    \caption{Angular distributions for the determination of the ORF from the \nuc{14}{C}$(p,d)$\nuc{13}{C} reaction at 35 MeV. Shown are curves for the angular distributions of the summed cross sections in the numerator (solid) and the denominator (dashed) of Eq. (\ref{eq-rstr}) and for the 0.0 MeV (short dashed), 3.089 MeV (dotted), 3.685 MeV (dash-dotted), and 3.854 MeV (double-dashed) MeV states of \nuc{13}{C}.}
    \label{fig-fig2}
\end{figure}

With the experimental SFs extracted for each state, we apply Eq. (\ref{eq-rstr}) to calculate the ORF for \nuc{14}{C} from the \nuc{14}{C}$(p,d)$\nuc{13}{C} reaction. The details are shown in Fig. \ref{fig-fig2}, where the individual contributions of the terms in the numerator of the right hand side of Eq.  (\ref{eq-rstr}) are shown as short dashed, dotted, dash-dotted, and double-dashed curves and their summed experimental angular distributions are plotted as the solid curve. The corresponding individual contributions in the denominator are not shown for the sake of clarity. Their sums are shown as the dashed curve after being normalized to the summed experimental cross sections at $\theta_\textrm{c.m.}=12$ degrees. This results in an ORF of $R_s^\textrm{tr}=0.565\pm 0.048$. Our previous statement that the contributions of the two positive-parity states to the ORF are negligible becomes obvious in this figure. The uncertainty of this $R_s$ value is evaluated with Eq. (\ref{eq-rs-err}).

\begin{figure}[htbp]
    \centering
    \includegraphics[width=0.48\textwidth]{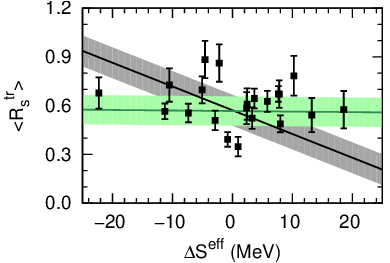}
    \caption{Averaged ORFs from the reactions analyzed in this work (squares). The green and grey bars represent the linear dependence of the ORFs on the $\Delta S^\textrm{eff}$ values fitted assuming a free or a fixed slope, respectively. See the text for details.}
    \label{fig3}
\end{figure}

Similar analysis have been made for all the other reactions. For cases when experimental data of a reaction are available at several incident energies, we calculate the \textit{averaged} ORF, $\langle R_s^\textrm{tr}\rangle$. The resulting averaged ORFs are plotted in Fig. \ref{fig3} against their corresponding effective isospin asymmetry $\Delta S^\textrm{eff}=S_{n}-S_{p}+\bar{E}_f$ \cite{Simpson-PRC-2009}, where $\bar{E}_f$ is the effective final state excitation energy, which is, similar as that defined in Refs. \cite{Simpson-PRC-2009, Tostevin-PRC-2014} for knockout reactions, an average of the excitation energies of each state weighted by the corresponding integrated transfer cross sections.

Although obtained from the consistent three-body model reaction methodology with which all reactions are analyzed with the same procedure without free parameters, the ORFs and their averaged values still scatter considerably in the $\langle R_s^\textrm{tr}\rangle$-$\Delta S^\textrm{eff}$ plot. Results of the systematic analysis of knockout reactions in Ref. \cite{Tostevin-PRC-2014} suggest that the RFs $R_s^\textrm{ko}$ depend linearly on $\Delta S^\textrm{eff}$: $R_s^\textrm{ko}=-1.46\times10^{-2}\times\Delta S^\textrm{eff}+0.596$ (this is obtained by fitting the data in Fig. 1 of Ref. \cite{Tostevin-PRC-2014}). We also assume linear dependence of ORFs on the $\Delta S$ values in transfer reactions: $R_s^\textrm{tr}(\Delta S^\textrm{eff})=a\times\Delta S^\textrm{eff}+b$. The slope $a$ and the parameter $b$ are obtained by least square fitting of the scattered data in Fig. \ref{fig3}. These fittings are made with (i) assuming $a=-1.46\times 10^{-2}$ MeV$^{-1}$, which is the same as the slope in the systematics of knockout reactions, and letting $b$ to vary freely, and (ii) letting both $a$ and $b$ to vary freely. The results are shown in Table. \ref{tab-chi2} and are visualized as shaded bars in Fig.  \ref{fig3}. The widths of these bars, listed in the last column of Table. \ref{tab-chi2}, represent the standard deviations of the distances of the scattered points from the fitted lines. The fact that the $\chi^2$ value associated with the $a=-3.37\times10^{-4}$ MeV$^{-1}$ case is considerably smaller than with the $a=-0.0146$ MeV$^{-1}$ case suggests that the ORFs are nearly independent on the isospin asymmetry. This value of $a$ is very close to those obtained in Ref. \cite{Moro-PLB-2018} for $(p,2p)/(p,pn)$ reactions with the Paris-Hamburg potentials. A bootstrap analysis \cite{bootstrap-sa-1983,CH89-pr-1991,Pang-PRC-2009} of the slope, which was made with 1000 times resampling of the $\langle R_s^\textrm{tr}\rangle$ values in the linear least square fittings resulted in an averaged value of $a$ and its standard deviation being $\langle{a}\rangle =8.55\times10^{-5}$ MeV$^{-1}$ and $\sigma_a=2.68\times10^{-3}$ MeV$^{-1}$, respectively. This result further suggests that the ORFs are independent on the isospin asymmetry.

\begin{table}[htbp]
    \centering
    \caption{Parameters of the linear fitting of the data in Fig. \ref{fig3} and their corresponding $\chi^2$ per degree of freedom. The last column is for the standard derivations of the distances between the points and the fitted lines.}
    \begin{tabular}{rrrrr}
        \hline
        slope & \multicolumn{1}{c}{$a$ (MeV$^{-1}$)} & \multicolumn{1}{c}{$b$} & \multicolumn{1}{c}{$\chi^2$} & \multicolumn{1}{c}{$\Delta_{R_s}$} \\\hline
        fixed & $-1.46\times10^{-2}$  & 0.572 & 5.9 & 0.0944 \\
        free  & $-3.37\times10^{-4}$  & 0.567 & 3.4 & 0.0896 \\\hline
    \end{tabular}%
    \label{tab-chi2}%
\end{table}%

\section{Summary}

In summary, to understand the reduction of the single-particle strength and its dependence on the isospin asymmetry is an important subject in nuclear physics. The RFs that were used to be referred to are found to be defined differently in transfer and knockout reactions. The RFs defined in knockout reactions involve all the bound excited states of the reaction residues but those in transfer reactions are for each single particle state. We define an overall reduction factor for the analysis of $(p,d)$ reactions, which include contributions from, in principle, all the probed bound states of the residual nuclei. This permits a proper comparison between the RFs extracted from transfer and knockout reactions. The ORFs extracted from a systematic analysis of 103 sets of angular distributions of $(p,d)$ reactions on 21 even-even nuclei with atomic mass numbers ranging form 8 to 56 with a consistent three-body model reaction methodology are found to have nearly no isospin dependence over a wide range of $\Delta S$ values. This is consistent with the recent measurement of $(d,t)$, $(d,\nuc{3}{He})$, $(p,2p)$, and $(p,pn)$ reactions on nitrogen and oxygen isotopes and \textit{ab initio} calculations \cite{Barbieri-IJMPA-2009, Flavigny-PRL-2013, Kay-PRL-2013, Atar-PRL-2018, Kawase-PTEP-2018, Moro-PLB-2018, Flavigny-PRC-2018}. Our result suggests that the known systematical discrepancy in the dependence of the RFs on the isospin asymmetry obtained from transfer and knockout reactions is not due to the exclusive or inclusive treatment of the experimental data. It is worthy to add at this place that although nuclear structure theories always endeavor to describe the properties of nuclei as precisely as possible, sometimes some overall comparisons between experimental and theoretical results may also be valuable. The ORF inducted in this work for the analysis of transfer reactions would be useful for such purposes.

\section*{Acknowledgments}
This work is supported by the National Natural Science Foundation of China (Grants Nos.  U1432247, 11775013, 11775004, and 11775316) and the national key research and development program (2016YFA0400502). D.Y.P. thanks Prof. J.A. Tostevin for stimulating communications in the Summer of 2015.

\section*{References}
\bibliography{pd-reduction-factor}

\end{document}